# Exchange bias in a core/shell magnetic nanoparticle: Monte Carlo simulation

M. H. Wu, Q. C. Li, and J. M. Liu[a]

*Laboratory of Solid State Microstructures, Nanjing University, Nanjing 210093, China*

*International Center for Materials Physics, Chinese Academy of Sciences, Shenyang, China*

[Abstract] By using Monte Carlo simulation on a ferromagnetic core/antiferromagnetic shell nanoparticle, we investigate in details the exchange bias of the magnetic hysteresis as a function of both core radius and shell thickness, at low temperature. It is found that the exchange bias is very sensitive to the core radius and a small variation of the radius may lead to a big fluctuation of the bias. In a general tendency the exchange bias is enhanced by increasing shell thickness and decreasing core radius. The intrinsic correlation between the exchange bias and the spin configuration on the core-shell interface is demonstrated. We further investigate the dependence of the exchange bias on temperature and random field inside the nanoparticle, indicating a monotonous decreasing of the bias with the magnitude of random field and temperature, respectively.

PACS numbers:    75.75.+a, 05.10.Ln, 75.50.Tt

Keywords:    core/shell nanoparticle, exchange bias, Monte Carlo simulation

[a] Corresponding author, E-mail: liujm@nju.edu.cn

## I. Introduction

When a ferromagnetic (FM) component is coherently contacted with an antiferromagnetic (AFM) one, and submitted into a continuously decreasing temperature $T$ under an external magnetic field $H$ through the Neel point $T_N$ of the AFM component, where $T_N$ is usually lower than the Curie point $T_C$ of the FM component, a shift of the hysteresis loop along the $H$-axis will be observed. This effect is known as the exchange bias (EB) and represents an essential feature in FM/AFM composite structure and is of special significance for magnetic recording applications [1]. The EB effect was first discovered in partially oxidized Co particles decades ago [2]. Up to date there have been hundreds of materials combinations reported in order to achieve optimal performances for practical application purposes [3].

Recent interest comes to the EB effect in magnetic nanostructures because of the promising high-density magnetic recording and memories, such as in giant magnetoresistive spin-valves and vertical recording geometry. For a review on this topic, readers may refer to Ref.[4]. A huge of theoretical and experimental works on the EB effect as a function of component dimension [5,6], temperature [7,8], or interface roughness [9,10] in layered FM/AFM structure were reported. It is worthy of mention that the core/shell nanoparticles as potential magnetic recording media are of interest, where usually the core is FM and the shell is AFM. In fact, in the earliest work on Co particles, the AFM CoO shell surrounding the FM Co core constitutes the core/shell nanoparticle. In such case, the well-ordered nanoparticle assembly constitutes a high-density recording plateform. In magnetic core-shell structure, a significant EB effect can be predicted because of the coherent interface coupling between the core and shell. Although in nanoparticle systems less experimental techniques can be applied for characterization of structure and property [11,12], many works on microscopic models, especially Monte Carlo (MC) simulations [13,14], including those partially addressing the EB phenomenology in nanostructures [15], have been presented. However, in literature on the MC simulations, a direct correlation between the EB and delicate spin configuration on the core/shell interfaces is not very clear, since the interfacial spin configurations can be very complicated, to be shown below.

It has been addressed that uncompensated interfacial spins play a crucial role on the EB [16-19], which is believed to depend mainly on the interfacial spin configuration and less on the particle size [20]. Therefore it is likely to explain those dependences by inspecting the interfacial spin configurations. It is interesting to note that the spin configuration on the core/shell interface is associated with the core radius in a fascinating manner. In the present paper, by a simple MC simulation on a cubic model lattice constituting a core/shell nanoparticle, we give a detailed investigation on the correlation of the EB with the spin configuration on the core/shell interface by varying the core/shell dimensions, random field and temperature.

## II. Model and simulation

### A. Model

We consider a spherical nanoparticle made of a FM core surrounded by an AF shell in a simple cubic lattice. The spin interaction is described by the Heisenberg model. In the presence of an external field $H$ applied along the easy-axis direction (z-axis), the lattice Hamiltonian can be written as [21]:

$$H/k_B = -\sum_{<i,j>} J_{ij} s_i \cdot s_j - \sum_i k_i (s_{iz})^2 - h \sum_i s_{iz}, \quad (1)$$

where $<i,j>$ refers to the summation over the nearest-neighboring spin-pairs, $J_{ij}$ is the spin interaction energy which should be separately defined for the core, shell and core-shell interface, i.e. $J_{ij}=J_{Sh}$ for the shell, $J_{ij}=J_{Co}$ for the core and $J_{ij}=J_{Int}$ for the interface; $k_i$ ($k_{Sh}$, $k_{Co}$ and $k_{Int}$) is the magnetocrystalline anisotropy factor, $h=\mu H/k_B$ with $k_B$ the Boltzmann constant, $s_i$ is the spin moment at site $i$ with magnitude 1/2 and arbitrary orientation. The first term in Eq.(1) accounts for the nearest-neighbor spin exchange energy. Similar to earlier work [21], because $T_N$ of the AFM shell usually is lower than $T_C$ of the FM core, we set $J_{Sh}=-0.5J_C=-5K$. For simplicity, the exchange for interfacial spins $J_{Int}=\pm J_{Sh}$ is assumed over the whole simulation, giving $J_{Int}=J_{Sh}$ for the AFM exchange and $J_{Int}=-J_{Sh}$ for the FM exchange on the core/shell interface. It will be found below that the sign of $J_{Int}$ has no influence on the exchange bias of the particle although the spin configuration on the core/shell interface can be somewhat different. The second term in Eq.(1) takes into account of the magneto-crystalline anisotropy energy (hereafter, z-axis is assigned as the easy-axis) which is fixed to $k_{Co}=1K$ for the core and $k_{Sh}=10K$ for the shell, noting that $k_{Sh}>>k_{Co}$. The third term is the Zeeman energy in the presence of nonzero $h$. For the choice of these parameters, we adopt the values given in earlier work [21] so that a comparison between our work and Ref.[21] can be made.

### B. Procedure of simulation

We simulate the static magnetic hysteresis loop and equilibrium spin configuration of the core/shell nanoparticle under the adiabatic approximation at very low temperature, followed by a brief discussion on the effect of temperature $T$ in the last section. The simulation is performed based on the Metropolis algorithm. The trial step of the spin updating is a combination of three kinds of trial steps as described by Hinzke and Nowak [22]. Two simulation paths will be employed here. First, the system is cooled from a temperature lower than $T_C$ of the core and higher than $T_N$ of the shell, in equiv-step down to $T$=0.1K in the presence of $h=h_{FC}$=4K along the z-axis. Then at $T$=0.1K we cycle the magnetic field between $h$=4K to $h$=-4K and then back to $h$=4K in a step of $\delta h$=-0.1K by using 300 Monte Carlo steps per spin at each field to evaluate the static hysteresis. Due to the expected exchange bias, we define the half of the hysteresis from $h$=4K to $h$=-4K as descending branch (branch I) and the other half (from $h$=-4K to $h$=4K) as ascending branch (branch II), noting that the spin configurations on the two branches do not show symmetry around zero point.

Second, the system is cooled down to $T$=0.1K in the same way as above, and then is allowed to reach the equilibrium state under zero-field condition. In this case we care for the spin configuration on the core/shell interface, by which it is possible to evaluate the net magnetization of the *shell spins on the core/shell interface* (CSIS spins) at zero field.

### C. Evaluation of exchange bias

The first roadmap is a calculation from the simulated hysteresis. The EB $h_{eb}$ is defined as:

$$h_{eb1} = ( h_c^+ + h_c^- )/2 , \qquad (2)$$

where $h_c^+$ is the coercivity on the +$h$-axis and $h_c^-$ the coercivity on the -$h$-axis of the hysteresis.

Alternatively, $h_{eb}$ can be calculated by correlating with the spin configuration in the core/shell interface, referring to Ref.[21]:

$$h_{eb2} = -J_{Int}(M_{Int}^+ + M_{Int}^-)/2, \qquad (3)$$

where $M_{Int}^{\pm}$ is the net magnetization of the CSIS spins at the positive (negative) coercivities $h_C^{\pm}$ respectively, which can be evaluated from the simulated spin configurations at $h_C^{\pm}$. In details, we count all of the CSIS spins at $h=h_C^{\pm}$, and sum them along z-axis to obtain $M_{Int}^{\pm}$.

Thirdly, we can also define another EB value by evaluating the net magnetization of the CSIS spins, $M_{Int}$, from the equilibrium spin configuration of the nanoparticle under zero field:

$$h_{eb3} = -J_{Int} \cdot M_{Int}, \qquad (4)$$

In the following sections, we shall present in details our investigation on the dependence of $h_{eb}$ on a series of system parameters and core/shell dimensions. In a general sense, the evaluated $h_{eb1}$, $h_{eb2}$ and $h_{eb3}$ are quite well consistent from one and another.

### III. Results and discussion

*A. Core thickness dependence*

Fig.1(a) presents the evaluated $h_{eb1}$ and $h_{eb2}$ as a function of core radius $d_{Co}$ respectively for both $J_{Int}=J_{Sh}$ (squares and circles) and $J_{Int}=-J_{Sh}$ (up triangles and down triangles), given shell thickness $d_{Sh}=3$. In fact, as $d_{Sh}>3$, the results are the same as shown here. In Fig.1(b) is plotted $h_{eb3}$ as a function of $d_{Co}$ as $J_{Int}=-5K$ and the results for $J_{Int}=5K$ remain the same. Clearly, it is shown that the evaluated $h_{eb1}$, $h_{eb2}$ and $h_{eb3}$ are quite well consistent from one and another. From the good consistence of $h_{eb1}$ and $h_{eb2}$, one doesn't find significant influence of the sign of $J_{Int}$ ($=\pm J_{Sh}$) on the evaluated $h_{eb}$.

For the $h_{eb}$-$d_{Co}$ relation, it is observed that upon increasing $d_{Co}$, $h_{eb}$ shows remarkably oscillating behavior at small $d_{Co}$, and then tends to a stable value when $d_{Co}$ becomes very large. Although $h_{eb}$ gradually decreases with increasing $d_{Co}$ in a general tendency, it is very sensitive to $d_{Co}$ as $d_{Co}$ is small. For instance, $h_{eb}\sim-1.7K$ at $d_{Co}=4$, while it becomes $\sim-0.55K$ at $d_{Co}=5$. This high sensitivity and the oscillating behavior of $h_{eb}\sim d_{Co}$ relation are not favored from the point of view of practical applications. What should be addressed here is the exceptional case of $d_{Sh}=1$, where some CSIS spins remain naked (not bonded) and the others are fully bonded with six nearest neighbors (NN spins). In this case and $d_{Sh}=2$, the shell spins can no longer hold AFM order and the exchange bias will disappear, which will not considered in this work.

### B. Interfacial spin configurations at $J_{int}=-5K$

The above effects are well-known phenomena and the underlying physics can be understood by analyzing the interfacial spin configurations. In fact, the exchange bias is essentially attributed to the net magnetization of all CSIS spins and has no much to do with other shell spins themselves unless the thermally assisted spin reversal takes effect. To highlight our understanding of the exchange bias and also the oscillating behaviors of $h_{eb}$ as a function of $d_{Co}$, it will be helpful to analyze the configuration of the CSIS spins and their reversal sequence upon external field cycling. Here we focus on the case of $J_{Int}=-5K$ and for $J_{Int}=5K$ similar analysis can be done.

As $d_{Sh}>2$, all of the CSIS spins are bonded with six NN spins, of which the possible number of core spins is 1, 2, or 3. Because of the strong anisotropy for the shell spins ($k_{Sh}=10K$), all shell spins prefer to align along ±z-axis with small orientation fluctuations. The configuration of the CSIS spins will be determined by the minimization of Hamiltonian $H$. One can evaluate the minimal of $H$ by counting the effective field applied to the spins under consideration, where the effective field sums up the external field $h$ and the equivalent field imposed by the NN spins.

Take the case of $|h|\leq 4K$ ($h_{max}=\pm 4K$) and $|J_{Int}|=5K$ as an example. It is seen that only those CSIS spins with three NN core spins have possibility to flip. For those CSIS spins with two NN core spins (i.e. four NN shell spins), the equivalent field imposed by the four NN shell spins is $4\times 5K/2=10K$, while that imposed by the two NN core spins equals 5K. Since one has $10K>5K+|h|$, those CSIS spins with only two NN core spins are pinned and have no chance to reverse during the field cycle sequence, as long as $|h|<5K$ and no thermal fluctuations are counted. This analysis also applies to those CSIS spins with only one NN core spin if any. Therefore, the CSIS spins with one or two NN core spins have no response to the external field cycling and then no contribution to the hysteresis loop.

Note here that for core-shell structure, no CSIS spin can have more than three NN core spins. We only need to care for those CSIS spins each with three NN core spins. Furthermore, the three NN core spins must be in the same direction, and so do the three NN shell spins. Keeping this feature in mind we can classify all CSIS spins into two categories. If all the NN spins (including NN core spins and NN shell spins) of a CSIS spin are in the same direction, i.e. this spin is in opposite direction to all of its six NN spins, this CSIS spin is defined as category-B spin (B-spin). Otherwise, the three NN core spins must be in the opposite direction to the three NN shell spins, allowing the CSIS spin to be defined as category-A spin (A-spin). The A-spins have the same orientation with its three NN core spins.

We first explain the $h_{eb3} \sim d_{Co}$ relation as evaluated from the zero-field equilibrium configuration of the nanoparticle, as shown in Fig.1(b). It can be noted that $M_{Int}^{+} = M_{Int}^{-} = M_{Int}$, and $M_{Int}$ can be easily calculated. Here we take the case of $d_{Co}$=5 and $d_{Sh}$>2 as an example. Of all 290 CSIS spins, 162 spins align in the same direction and the other 128 spins are in the opposite direction. If $J_{Int}$=-5K, the equivalent field imposed by the NN core spins is always opposite to the external field $h$ during the external field cycling. This equivalent field outweighs the external field, enabling the net magnetization of all the CSIS spins in opposite direction to the external field and core spins. We obtain $M_{Int}$=(128-162)/290=-0.11, and $h_{eb}$=-0.55. Similarly, if $J_{Int}$=5K, we have $M_{Int}$=(162-128)/290=0.11 and $h_{eb}$=-0.55 too. In this way, we evaluate the exchange bias as a function of $d_{Co}$, as shown in Fig.1(b).

It is also useful to note $M_{Int}=\Delta N_{Int}/N_{int}$, where $\Delta N_{Int}$ is the difference between the numbers of the CSIS spins with two opposite directions and $N_{Int}$ the total number at zero field. We present $\Delta N_{Int}$ and $N_{Int}$ as a function of $d_{Co}$ respectively in Fig.2(a) and then $M_{Int}$ as a function of $d_{Co}$ in Fig.2(b). Clearly, one has $N_{Int} \sim d_{Co}^2$ while $\Delta N_{Int}$ oscillates with increasing $d_{Co}$, resulting in an oscillating decreasing behavior of $M_{Int}$ as a function of $d_{Co}$, thus the oscillating behavior of $h_{eb3}$ against $d_{Co}$. However, perfect spherical particles are not available in experiment, therefore in experiment, oscillation is eliminated as a result and only the decreasing behavior is observed.

Secondly, we come to explain the $h_{eb2} \sim d_{Co}$ relation as shown in Fig.1(a). Upon external field cycling, those CSIS spins with three NN core spins may reverse, which may influence on the EB. The reversal sequence can be identified from the hysteresis of $M_{Int}$ for the CSIS spins with three NN core spins against external field $h$. We present the evaluated $M_{Int}$~$h$ hysteresis loops at different $d_{Co}$ in Fig.3(a) for $J_{Int}$=-5K. It is seen that the $M_{Int}$~$h$ hysteresis upon different $d_{Co}$ are very different from one and another, predicting the remarkable fluctuations of the core/shell interface coupling at different $d_{Co}$, which are intrinsically responsible for the oscillating pattern of $h_{eb}$~$d_{Co}$ relation.

We observe the intrinsic relationship between the bias and the pattern of the $M_{Int}$~$h$ hysteresis at $J_{Int}$=-5K. If the up- or down-apex of the $M_{Int}$~$h$ hysteresis is remarkable, the exchange bias will be small. For details, one may take the $M_{Int}$~$h$ hysteresis at $d_{Co}$=9 as an example. We first consider branch I of the hysteresis. At the beginning when $h$>0, the exchange energy for the A-spins is very small and a small external field $h$ can force them to reverse if they are initially in an opposite direction to $h$. However, the B-spins are highly stable against opposite external field $h$ unless $h$ is very large. For the B-spins aligned in parallel to the NN core spins, once $h$ decreases down to a negative value at which the NN core spin reversal occurs, they will turn into A-spins and at the same time those A-spins will turn into B-spins. This sequence can be schematically shown in Fig.4 from step 1 to step 2. As $d_{Co}$=9, there are total 918 CSIS spins with 384 spins align in the same direction and the other 534 spins are in the opposite direction to external field $h$. Therefore, we have $M_{Int}$=(384-534)/918=-0.163, and $h_{eb}$=-$J_{Int}$·$M_{Int}$=-0.82. When $h$ reduces down to zero and becomes negative, 48 A-spins among the total 120 CSIS spins with three NN core spins will reverse, leading to a decreasing of $M_{Int}$ by 48×2/918=0.105, so $M_{Int}$=-0.163-0.105=-0.268. Furthermore, when $h$ becomes even more negative ($h$~ -3K) so that the core spins' reversal occurs, the 48 A-spins will reverse back to their initial state, so $M_{Int}$ comes back to -0.163 again. This sequence leads to the appearance of a local down-apex in the $M_{Int}$~$h$ hysteresis, as shown in Fig.3(b).

However, for branch II of the hysteresis (step 3 to step 4 in Fig.4), the core spins on the core/shell interface should be able to reverse earlier than other core spins [15] because of the negative exchange bias. If the core spins on the interface can reverse in prior to the A-spins, these A-spins will be unable to reverse, and thus the apex will disappear, as shown in Fig.3(b) too. Otherwise, one can observe an up-apex in the hysteresis, like the case of $d_{Co}$=20 as shown in Fig.3(b). Similar case applies to $d_{Co}$=15, 17, and 23, as identified in Fig.3(a). In case of few

A-spins and B-spins, such as for nanoparticles of small core radius $d_{Co}$=3, 5 or 7, we observe no apex in both branches I and II, as shown in Fig.3(a).

Sometimes the effect of these apexes on exchange bias is considerable. Take $d_{Co}$=9 stated above for instance, when the apex that appears is taken into account, we obtained -$h_{eb2}$ =1.07, which is larger than -$h_{eb3}$ =0.82 obtained without considering the apex. Besides, both up-apex and down-apex contribute to the enlargement of coercivity. According to Ref.21, $h_c = h_c^0 + J_{Int}( M_{Int}^+ - M_{Int}^-)/2$, and both apexes can enlarge $M_{Int}^+ - M_{Int}^-$. Taking the example of $d_{Co}$=9, its coercivity is increased by $J_{Int}( M_{Int}^+ - M_{Int}^-)/2$=0.505 as a result of the apex.

In addition, there is another factor contributing to the difference between $h_{eb2}$ and $h_{eb3}$. As $d_{Co}$=6 or 12, because of certain configuration, $M_{Int}$ is so small that Zeeman energy acting on the surface spins of the AFM can overcome the AFM-FM exchange interaction. According to Ref.23, exchange bias will become positive then.

### C. Interfacial spin configurations at $J_{Int}$=5K

Although the hysteresis for the whole lattice at $J_{Int}$=5K remains quite similar to that at $J_{Int}$=-5K, given the same values for other parameters, the mechanism for the EB is different, which can be understood by analyzing the $M_{Int}\sim h$ hysteresis. At the beginning of branch I, the number of B-spins, $M_{Int}$, and magnetization of the core $M_{Co}$, are all in the same direction with $h$. Analogically, when $h$ decreases from the positive maximal $h_{max}$=4K (branch I), all B-spins align with the external field, which will reverse as $h$ reduces down to zero and becomes negative. This will lead to a decreasing step in $M_{Int}$, as shown in Fig.3(c) at $d_{Co}$=9 and 10. Meanwhile, all the A-spins remain in the same direction as core spins. As long as $h$ further decreases down to a negative value at which the NN core spins begin to reverse, the earlier B-spins will turn into A-spins and vice verse. Consequently, a tremendous drop step in $M_{Int}$ will be expected, accompanying with transfer of the earlier A-spins into B-spins. Similar sequence will occur when $h$ increases from the negative maximal –$h_{max}$=-4K (branch II), which leads to two increasing steps in branch II of the $M_{Int}\sim h$ hysteresis.

However, it is observed in Fig.3(c) that as $d_{Co}$=4 and 7, we observe only one decreasing step on branch I and one increasing step on branch II. The reason lies in that for the two cases the number of B-spins is very limited so that the other step is too small to be identified. In general, for any $d_{Co}$, if the number of B-spins is comparable to that of A-spins, one will observe two steps on each branch of the hysteresis, otherwise only one step on each branch can be observable.

### D. Shell thickness dependence

In this section, we come back to study $h_{eb}$ as a function of $d_{Sh}$. It is easily understood that the dependence of $h_{eb}$ on $d_{Sh}$ makes sense only at $d_{Sh}$=1, 2 and 3, beyond which no dependence of $h_{eb}$ on $d_{sh}$ is expected. The simulated results are presented in Fig.5(a) and (b) for $J_{Int}$=-5K and 5K, respectively. At $d_{Sh}$=1, most of the CSIS spins are partially naked with only 4 or 5 NN spins. Since the AFM spin exchange energy is not large enough to resist against the interfacial spin interaction, almost all of the shell spins will reverse in association with the reverse of the core spins in response to cycling of external field $h$. Also because of very unstable AFM shell spins, one has $M_{Int}^{+} \sim -M_{Int}^{-}$, leading to $h_{eb}$~0, as shown in both Fig.5(a) and (b) for $d_{Sh}$=1.

Once $d_{Sh}$=2 and more, all the CSIS spins have six NN spins. If $J_{Int}$=-5K, referring to Fig.5(a), the hysteresis remains unchanged with increasing $d_{Sh}$ once $d_{Sh}$>1. For $J_{Int}$=5K, $h_{eb}$ no longer changes as $d_{Sh}$>2, as shown in Fig.5(b). This indicates that the shell spins (excluding those CSIS spins) are offered enough high stability against reversal of the core spins so long as $d_{Sh} \geq 2$. In such cases, the results presented in Sec.III-B and III-C make sense.

### E. Effect of random field

In real materials, the spin interaction can't be spatially homogeneous and always includes internal random field to some extent. We also investigate the effect of random field on the exchange bias. In this work, the random field applied to the lattice is assumed to follow the Gaussian distribution:

$$f(J) = \frac{1}{\sqrt{2\pi} j} e^{-\frac{(J-J_0)^2}{2j^2}}, \quad (4)$$

where $j$ is the variance of exchange interaction $J$, which should be separately defined for the core, shell and core-shell interface, i.e. $j=j_{Sh}$ for the shell, $j=j_{Co}$ for the core and $j=j_{Int}$ for the core/shell interface. The expectation of $J$, $J_0$, is set $J_0^{shell}$ =-5K for the shell, $J_0^{core}$=10K for the core, and $J_0^{int}$=-5K for the core/shell interface. We set $j_{Co}$:$j_{Int}$:$j_{Sh}$=|$J_0^{core}$|:|$J_0^{int}$|:|$J_0^{shell}$|=2:1:1, $d_{Co}$=9 and $d_{Sh}$=3 in our simulation as a demo. Because the variance of the sum of independent random variables equals the sum of each variable's variances, for a CSIS spin with three NN core spins, the variance of the equivalent field imposed by the 6 NN spins is $j_{NN}= \sqrt{3\times 1^2 + 3\times 0.5^2} j_{Co} \approx 1.94 j_{Co}$. For a CSIS spin with two NN core spins, $j_{NN}=\sqrt{2\times 1^2 + 4\times 0.5^2} j_{Co} \approx 1.73 j_{Co}$.

The simulated $M_{Int}$~$h$ hysteresis loops under different random field amplitude $j_{Co}$ are shown in Fig.6(a). As $j_{Co}$>0, the exchange energy of some A-spins becomes nonzero. When $h$ crosses over zero from positive value, those A-spins with positive exchange energy will reverse, while those with negative exchange energy may not, upon the magnitude of random field $j_{Co}$. As mentioned earlier in Sec.III-B, for $j_{Co}$=0, no apex on the hysteresis will be possible when $h$ increases from –$h_{max}$ (branch II), since the core spins will reverse in prior to the A-spins. However, as $j_{Co}$>0, some A-spins with positive exchange energy may reverse in prior to the core spin reverse, resulting in the apex in the hysteresis once more, although the apex is relatively diffusive. This effect is clearly shown in Fig.6(a), and it will be more significant when $j_{Co}$ is larger.

As $j_{Co}$ reaches up to 4.1K and more, for branch II, more and more A-spins have their exchange energy surpassing $j_{NN}$/2=1.94$j_{Co}$/2=4K. The equivalent field imposed by their NN spins outweighs the external field, which enables them to reverse to the positive direction before $h$ rises to zero and becomes positive. Meanwhile those with negative exchange energy remain not to reverse, acting as if there were no random field. As a result, $M_{Int}^+$ will rise and |$h_{eb}$| will decrease rapidly.

When $j_{Co}$ approaches to 10.4K and goes on increasing, more and more other shell spins, for example, those shell spins with two NN core spins as shown in Fig.2 become capable of reversing when the core spins begin to reversal on branch I, because their exchange energy exceeds ($j_{NN}$–10K)/2=(1.73$j_{Co}$–10K)/2= (18K-10K)/2=4K and outweighs the external field. At the same time, those with negative exchange energy remain stable. Thus $M_{Int}$ is greatly enhanced at the beginning of branch II, and $|h_{eb}|$ will continue decreasing as a result of rising $M_{Int}$. When $j$ attains to some certain value, $|h_{eb}|$ gradually stops declining.

### *F. Temperature dependence*

Finally, we simulate the temperature effect on the exchange bias, where temperature $T$ is included in the Metropolis algorithm of the simulation. The simulated $M_{Int} \sim h$ hysteresis loops under different $T$ for nanoparticle of $d_{Co}=9$, $d_{Sh}=3$ and $J_{Int}=-5K$ are shown in Fig.7(a). As $T$ arises, fluctuations of the curve become more and more considerable, while the antiferromagnetism of the shell diminishes. As $T$ approaches $T_N$, every shell spin receives less and less exchange energy from its NN shell spins to stabilize itself due to the declining antiferromagnetism. As a result, more CSIS spins will reverse simultaneously with the reversal of the core spins, which causes the shrinking difference between the absolute values of $M_{Int}^+$ and $M_{Int}^-$. At $T \approx T_N=2.0K$, where the AFM order of the shell disappears, all the CSIS spins receive zero exchange energy from its NN shell spins and are able to reverse once their NN core spins reverse, therefore $M_{Int}^+=-M_{Int}^-$, $h_{eb}=0$, as shown in Fig.7(b). This temperature is known as blocking temperature $T_B$, and our explanations above clarifies the phenomenon $T_B \approx T_N$ in high quality thin film systems with thick AFM layers [4].

### **IV. Conclusion**

In summary, through MC simulations based on the core/shell nanoparticle model, we have investigated in details the dependences of the exchange bias on both core radius and shell thickness. It has been demonstrated that the exchange bias is very sensitive to the core radius. A remarkable oscillating behavior of the bias with increasing core radius at small radius value has been identified, although in a rough sense the bias is higher when the core radius is smaller. A large bias is favored for a thick shell. Investigations on the interfacial spin configuration and its correlation with the EB enable us to understand the underlying physics. By similar methods we have investigated the dependence of the exchange bias on random field and temperature, and we find that exchange bias is reduced monotonously by the enhancement of random field and temperature.

**ACKNOWLEDGMENTS**

The authors thank the Natural Science Foundation of China (50332020, 10021001) and National Key Projects for Basic Research of China (2002CB613303, 2004CB619004).

**Figure Captions**

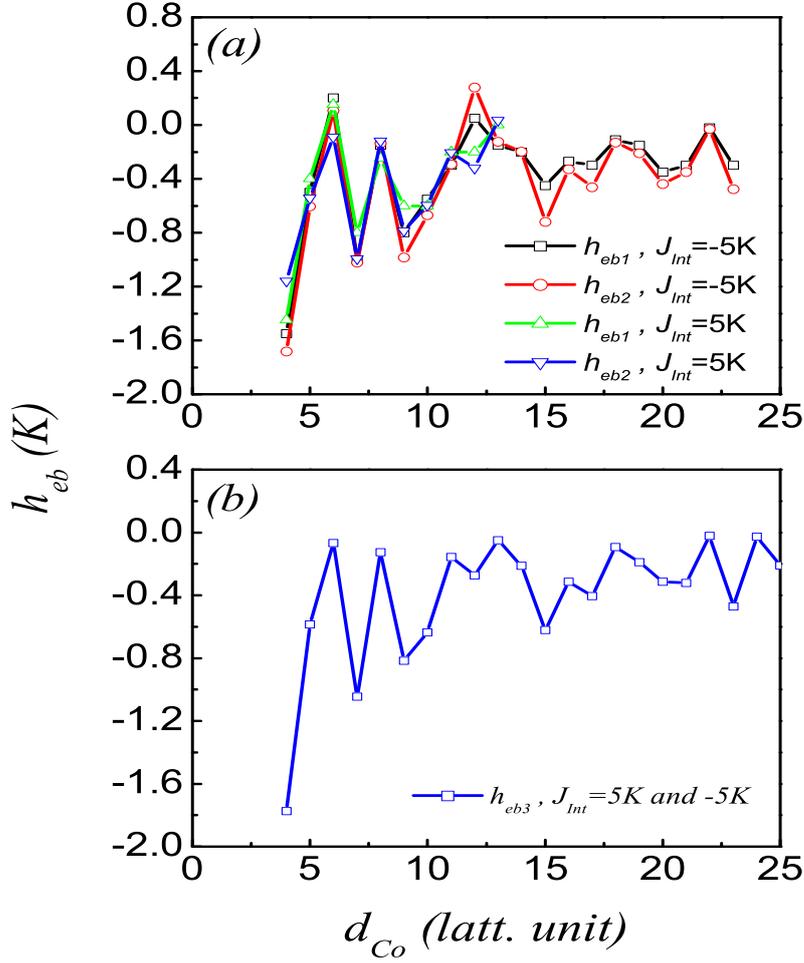

Fig.1. (color online) (a) Exchange bias $h_{eb}$ as a function of core radius $d_{Co}$. The squared dots and circles dots represent the data from $M_{Int}$ of the CSIS spins (Eq.(3)) and from the simulated hysteresis (Eq.(2)) as $J_{Int}=J_{Sh}$, respectively; and the up-triangle dots and down-triangle dots are for the data from $M_{Int}$ of the CSIS spins (Eq.(3)) and from the simulated hysteresis (Eq.(2)) as $J_{Int}=-J_{Sh}$, respectively. (b) $h_{eb}$ as a function of $d_{Co}$ as evaluated from Eq.(3) by counting $M_{Int}$ of the CSIS spins at zero field. $d_{Sh}=3$.

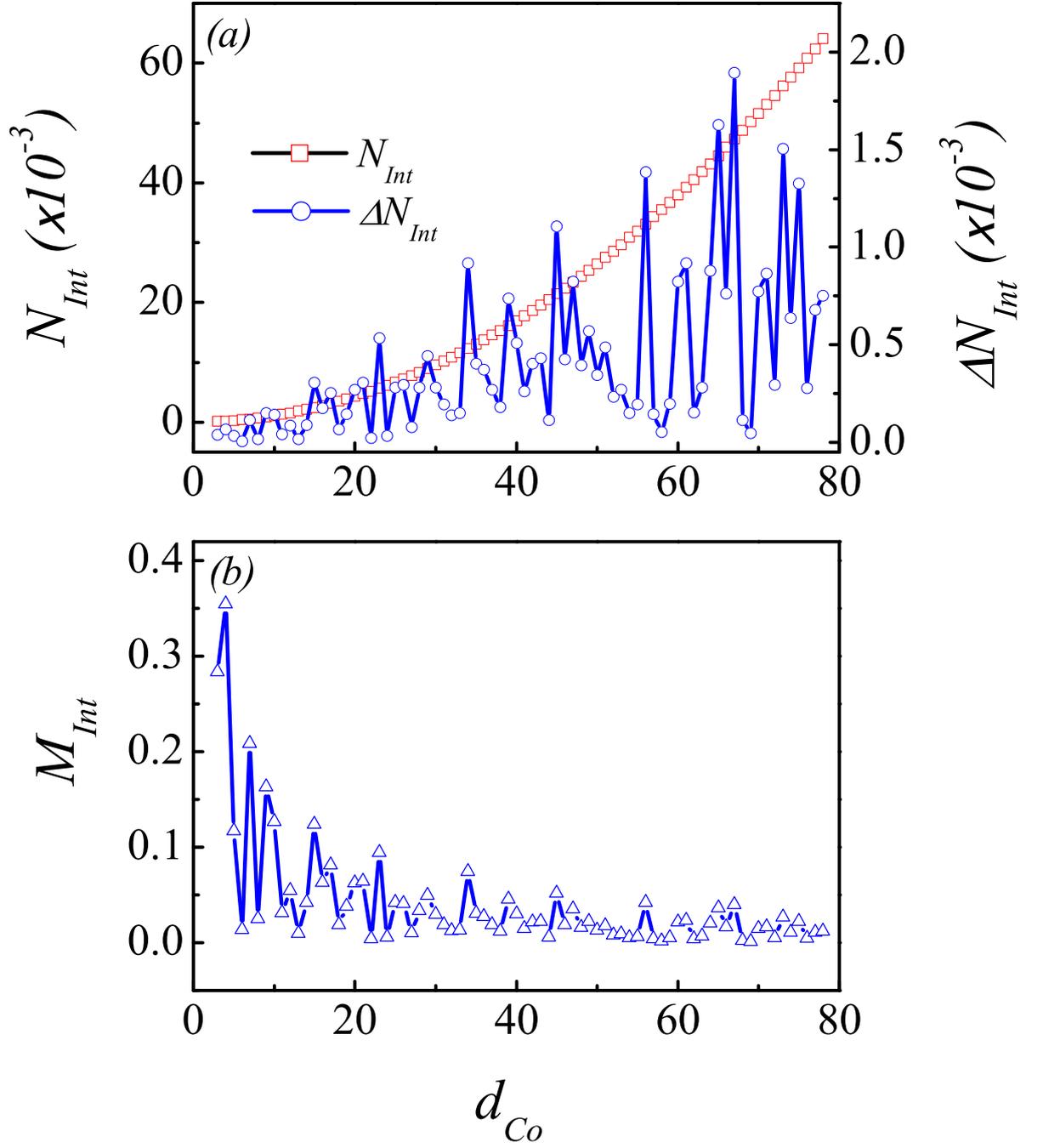

Fig.2. (color online) (a) Parameters $\Delta N_{Int}$ and $N_{Int}$ as a function of $d_{Co}$, respectively; (b) $M_{Int}$ as a function of $d_{Co}$. $J_{Int}$=5K and $d_{Sh}$=3.

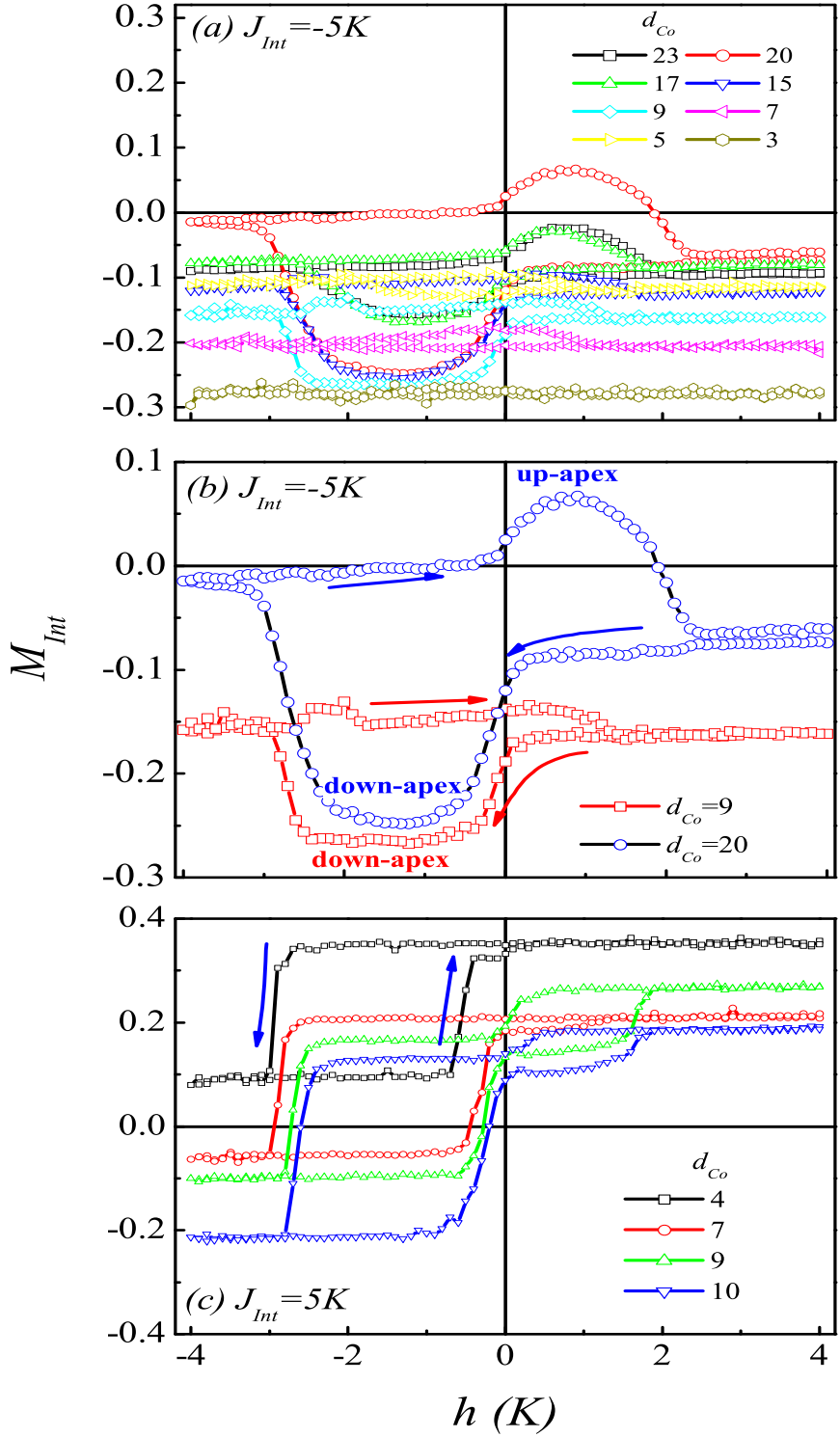

Fig.3. (color online) Simulated $M_{Int}$~$h$ hysteresis loops at $d_{Sh}=3$ and different $d_{Co}$ for (a) $J_{Int}=-J_{Sh}$ (b) two typical loops of $J_{Int}=-J_{Sh}$ and (c) $J_{Int}=J_{Sh}$.

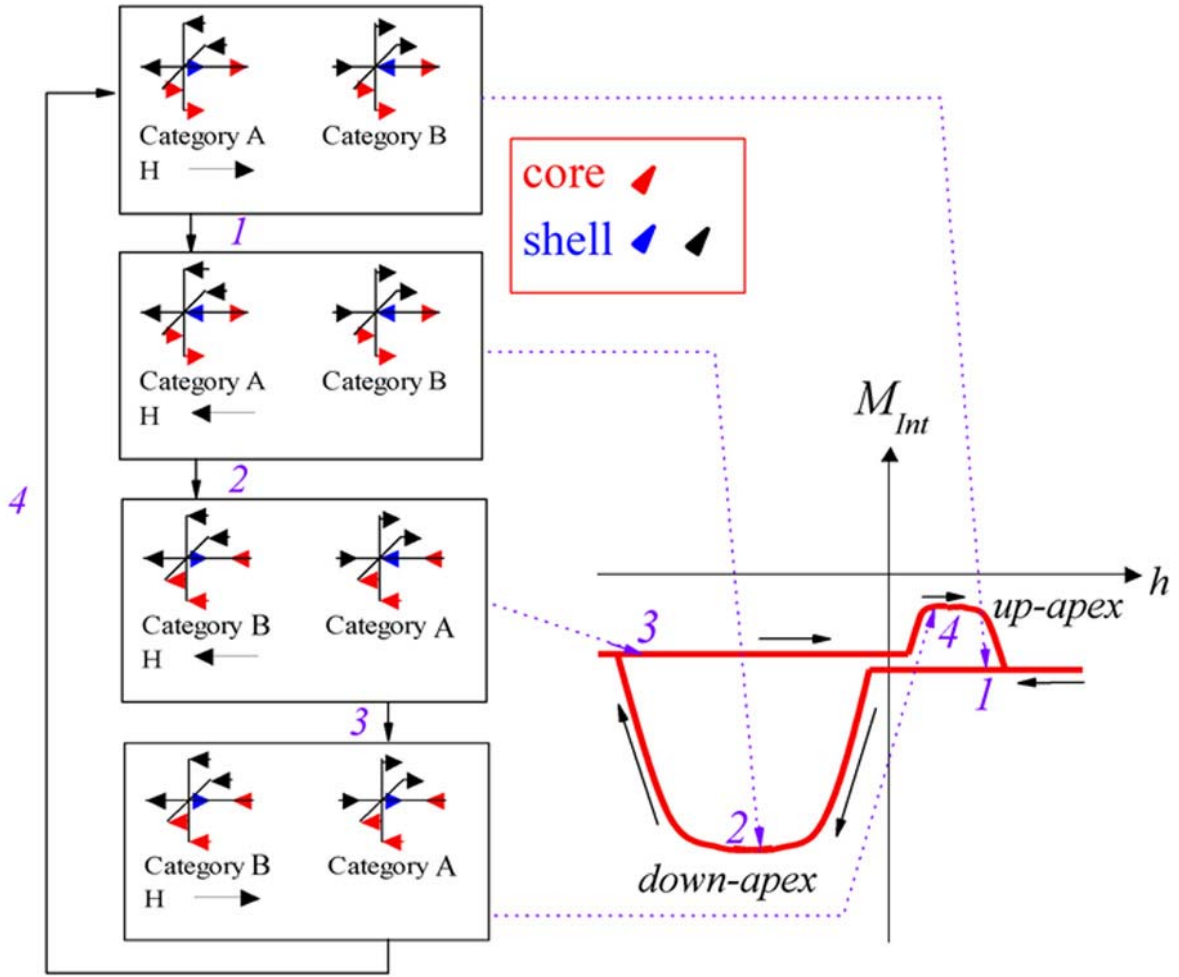

Fig.4. (color online) Schematic drawing of the alignments of a CSIS spin (blue) with its six NN spins (red for core spins and black for shell spins) at four different locations (1, 2, 3 and 4) of the $M_{Int}$~$h$ hysteresis. $J_{Int}$=-5K.

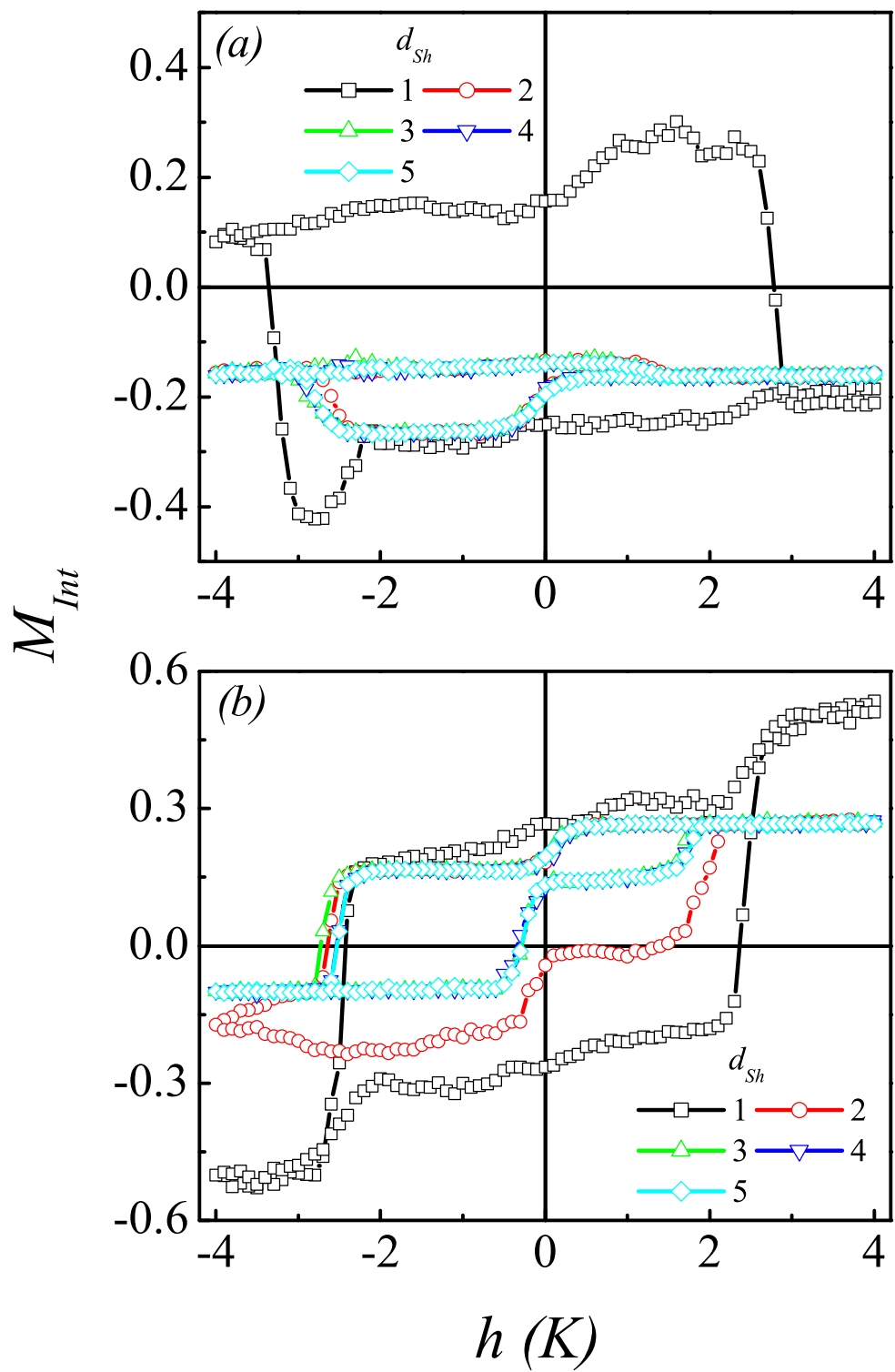

Fig.5. (color online) Simulated $M_{Int}$~$h$ hysteresis loops at $d_{Co}=9$ and different $d_{Sh}$ for both (a) $J_{Int}=-J_{Sh}$ and (b) $J_{Int}=J_{Sh}$.

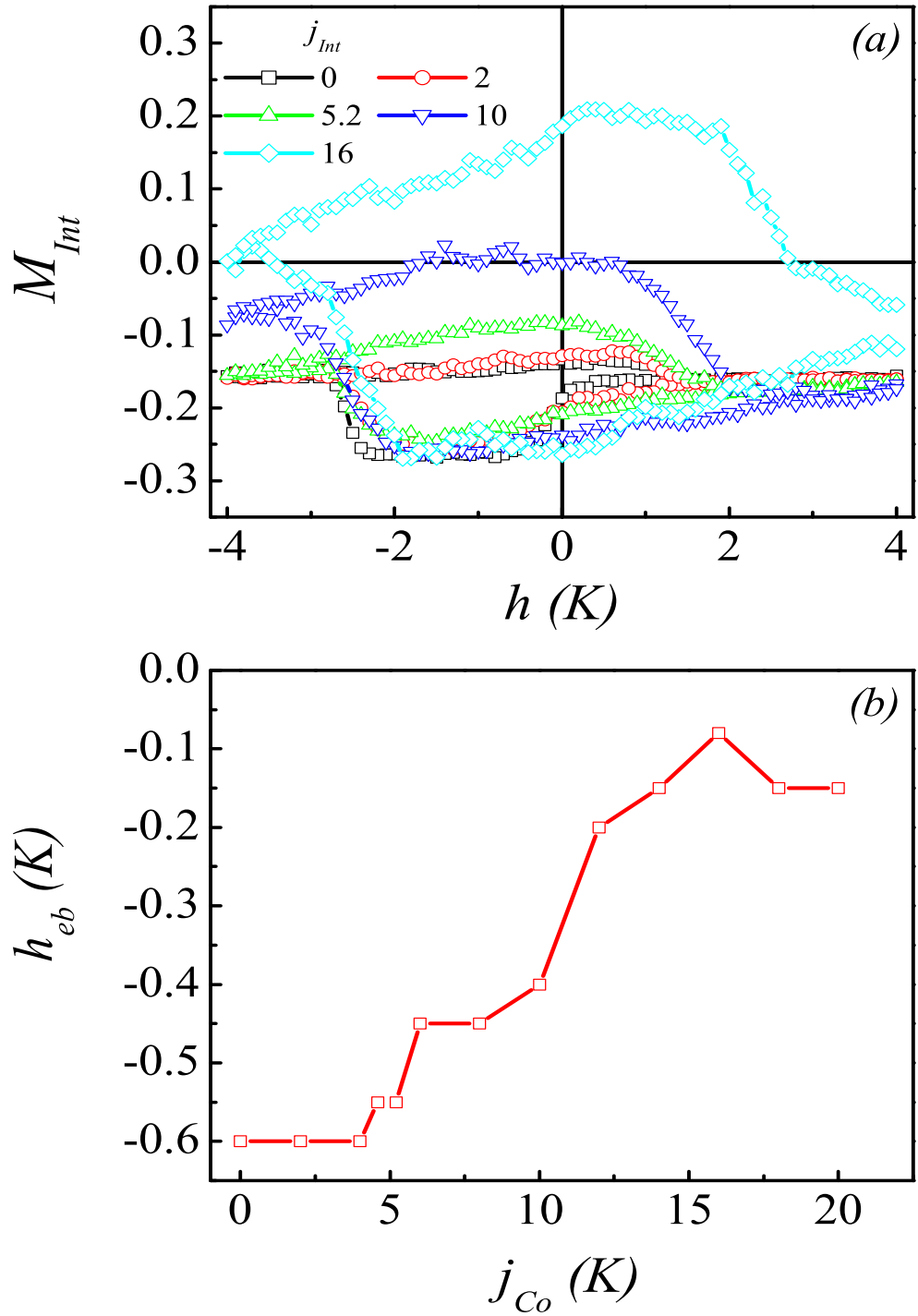

Fig.6. (color online) (a) Simulated $M_{Int}$~$h$ hysteresis loops under different $j_{Co}$. (b) Exchange bias $h_{eb}$ as a function of $j_{Co}$. $J_{Int}$=-5K, $d_{Co}$=9 and $d_{Sh}$=3.

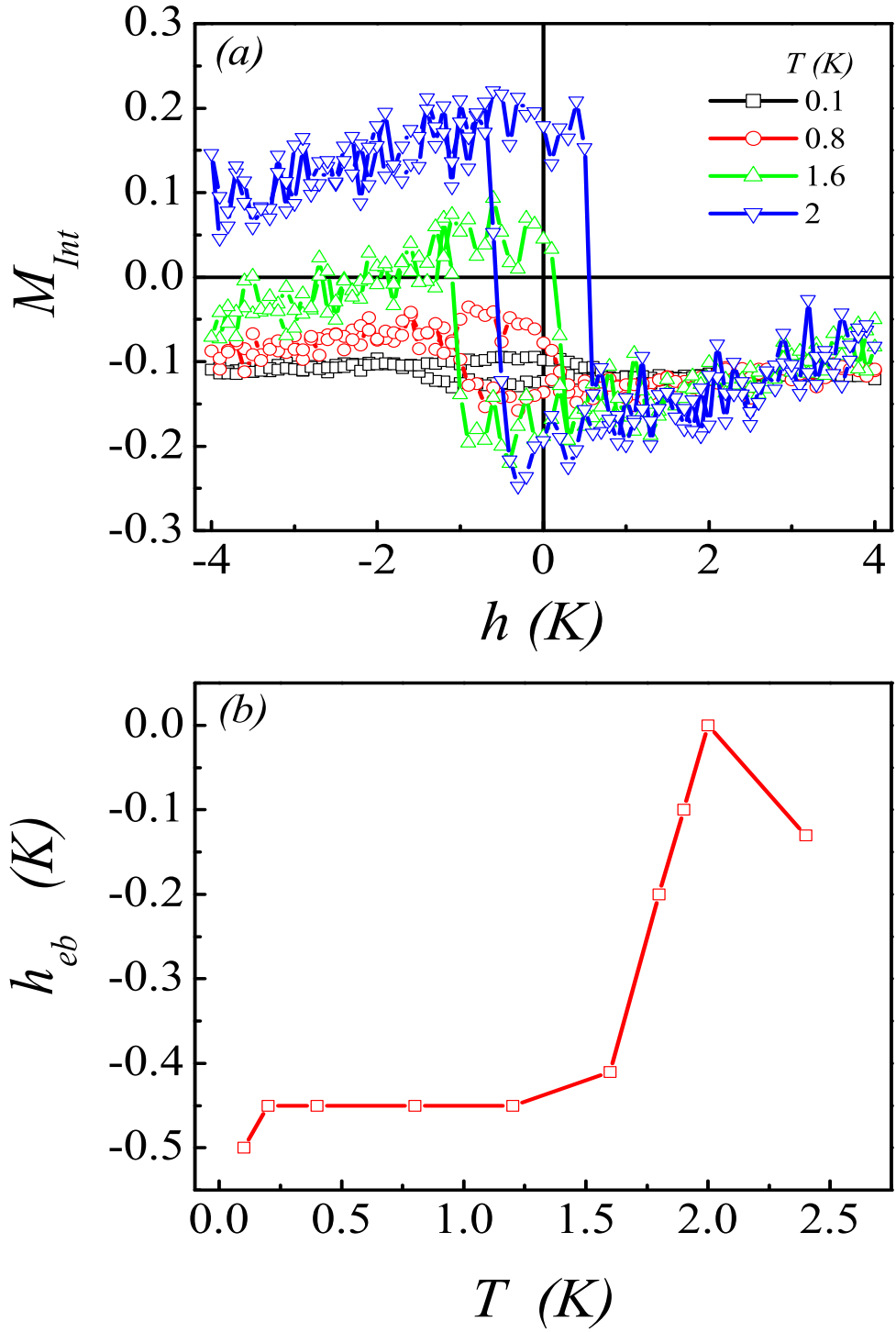

Fig.7. (color online) (a) Simulated $M_{Int}$~$h$ hysteresis loops under different temperature $T$. (b) Exchange bias $h_{eb}$ as a function of $T$. $J_{Int}$=-5K, $d_{Co}$=9 and $d_{Sh}$=3.